\documentclass{PoS}

\usepackage{amsmath}
\usepackage{dsfont}
\usepackage{booktabs}
\usepackage{graphicx}

\newcommand{\Op}{\mathcal{O}} 
\newcommand{\C}{\mathcal{C}} 
\newcommand{\eins}{\mathds{1}} 

\title{Nucleon form factors with dynamical twisted mass fermions }

\ShortTitle{Nucleon form factors}

\author{Constantia Alexandrou,\speaker{Tomasz Korzec}, Giannis Koutsou\\
        Department of Physics, University of Cyprus, P.O. Box 20537, 1678 Nicosia, Cyprus \\
        E-mail: \email{alexand@ucy.ac.cy}, \email{korzec@ucy.ac.cy}, \email{koutsou@ucy.ac.cy}}

\author{Mariane Brinet, Jaume Carbonell, Vincent Drach, Pierre-Antoine Harraud\\
        Laboratoire de Physique Subatomique et Cosmologie, 53 avenue des Martyrs, 38026 Grenoble, France\\
        E-mail: \email{mariane@lpsc.in2p3.fr}, \email{carbonel@lpsc.in2p3.fr}, \email{drach@lpsc.in2p3.fr}, \email{harraud@lpsc.in2p3.fr}}
	
\author{R\'emi Baron\\
        CEA, Centre de Saclay, IRFU/Service de Physique Nucl\'{e}aire, F-91191
        Gif-sur-Yvette, France\\
        E-mail: \email{remi.baron@cea.fr}}

\abstract{
\center{\includegraphics[width=0.25\linewidth]{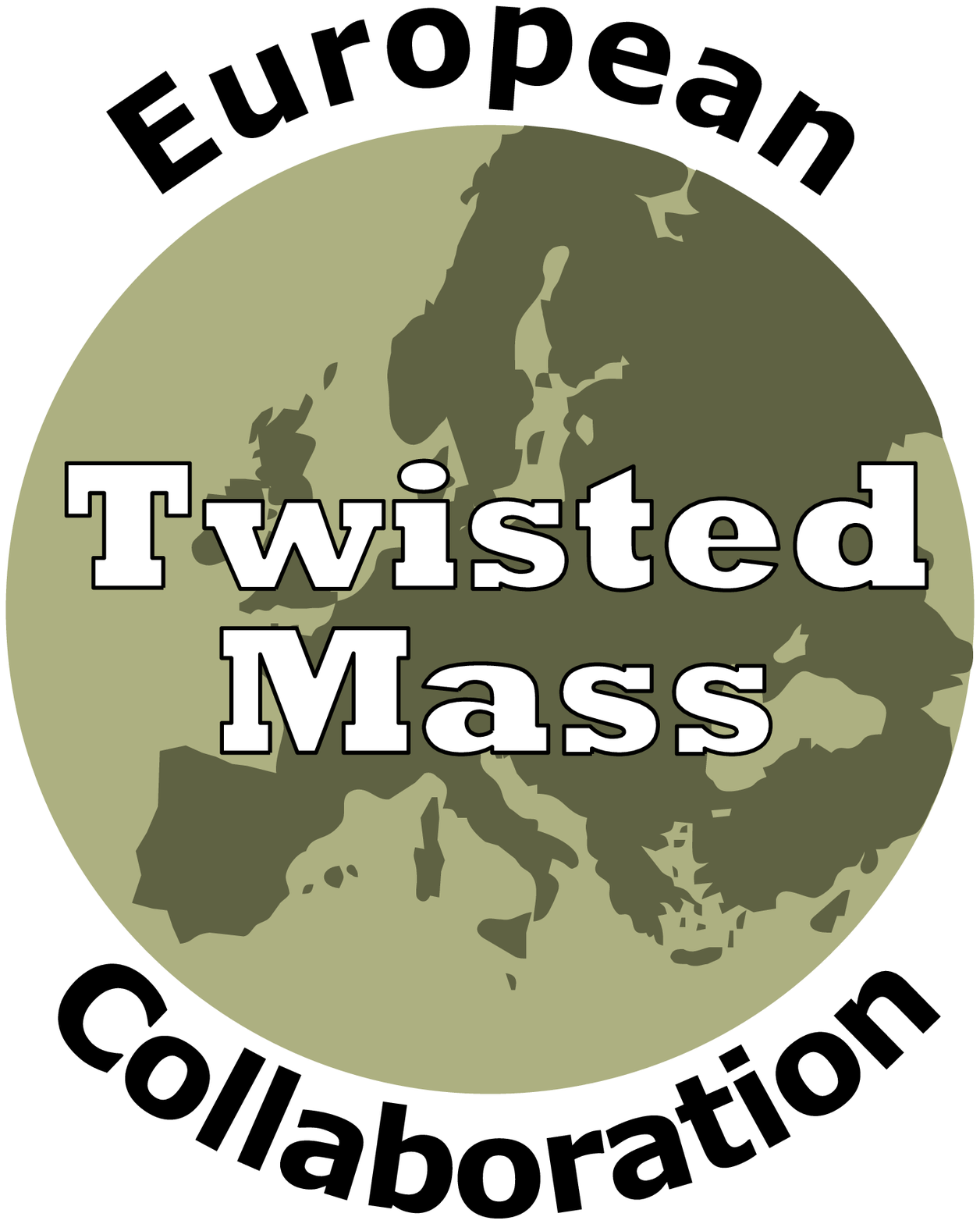}} 

\vspace*{1cm}
The electromagnetic and axial form factors of the nucleon
are evaluated  in twisted mass QCD with two degenerate flavors
of light, dynamical  quarks.
The axial charge $g_A$, magnetic
moment and the Dirac and Pauli radii
are determined for pion masses
in the range 
300 MeV to 500 MeV. 
}
\FullConference{The XXVI International Symposium on Lattice Field Theory \\
                 July 14 - 19, 2008\\
                 Williamsburg, Virginia, USA}

\begin{document}

\section{Introduction}
Recent calculations  in both 
the meson~\cite{Boucaud:2007uk,Jansen:2008wv} and baryon~\cite{Alexandrou:2008tn} sectors
within the twisted mass
formulation of lattice QCD~\cite{Frezzotti:2000nk} have yielded accurate results
on a wide range of observables of immediate relevance 
to experiment and phenomenology~\cite{Leutwyler:2007}.
This is a consequence of the fact that the
 simulations of the European Twisted Mass 
Collaboration (ETMC) cover a well chosen range of parameters, allowing
controlled continuum and chiral extrapolations and thus making 
a reliable connection to the physical regime possible.
 In the present work we apply the twisted mass
framework to the calculation of nucleon form 
factors.

Electromagnetic and axial form factors of the proton and the neutron 
are fundamental quantities that yield  information on their internal structure
such as their 
size, magnetization and axial charge. They
have been studied experimentally for a long time with steadily increasing 
precision, the latest generation of experiments reaching relative
precisions of down to about 1\%~\cite{Punjabi:2005wq}. 
The first lattice calculations
using dynamical fermions appeared only recently~\cite{Alexandrou:2006ru,Gockeler:2007ir,Hagler:2007xi} and the
available pion masses were rather large, often above 400 MeV.

Our goal is to calculate matrix elements of the form
$\langle N(p_f,s_f)| \Op_\mu| N(p_i,s_i)\rangle$, where
$|N(p,s)\rangle$ is the nucleon ground state, with the nucleon 
having momentum $p$ and spin $s$. $\Op_\mu$ is either the 
electromagnetic or the axial current. Using, in addition,
the pseudoscalar current will enable us to check phenomenological
consequences of chiral symmetry such as the Goldberger-Treiman 
relation~\cite{Alexandrou:2007zz}. 
The electromagnetic matrix element
can be expressed in terms of two Lorentz invariant form factors
that depend on the momentum transfer squared only. In Euclidean 
space-time the decomposition is
\begin{equation}
   \langle N(p_f,s_f)| V_\mu(0) | N(p_i, s_i) \rangle = \bar u(p_f,s_f) \left[\gamma_\mu F_1(Q^2)+\frac{\sigma_{\mu\nu}Q_\nu}{2m} F_2(Q^2) \right]u(p_i,s_i) \, ,
\end{equation}
where $Q=p_f-p_i$, $\sigma_{\mu\nu}=i[\gamma_\mu,\gamma_\nu]/2$, $m$ is the proton mass and
$u(p,s)$ is a solution to the free Dirac equation with mass $m$.
Instead of the Dirac and Pauli form factors $F_1$ and $F_2$   
the matrix element can be expressed in terms of the electric and magnetic Sachs form factors
\begin{equation}
   G_E = F_1(Q^2) - \frac{Q^2}{4m^2}F_2(Q^2) \, , \qquad \qquad
   G_M = F_1(Q^2) + F_2(Q^2) \, .
\end{equation}
Similarly the axial current matrix element can be written in terms of
the form factors $G_A$ and $G_p$,
\begin{equation}
   \langle N(p_f,s_f)| A_\mu(0) | N(p_i, s_i) \rangle = \bar u(p_f,s_f) \left[\gamma_5\gamma_\mu G_A(Q^2)+\frac{i Q_\mu \gamma_5}{2m} G_p(Q^2) \right]u(p_i,s_i) \, .
\end{equation}

\section{Calculation details}
\subsection{Wilson twisted mass QCD}
For our calculation we use a Wilson twisted mass fermion
 action (tmQCD)
at maximal twist angle 
together with a tree-level Symanzik improved gauge action.
More precisely, the fermion action is given by
\begin{equation}\label{TMaction}
   S^f = a^4\sum_x \bar\chi\left[\frac{1}{2}\left(\gamma_\mu(\nabla_\mu+\nabla^*_\mu)
                                 -a\nabla_\mu\nabla^*_\mu \right)
				 +m_{\rm crit} +i\gamma_5\tau_3\mu \right] \chi \, ,
\end{equation}
with lattice spacing $a$, the covariant forward and backward lattice derivatives
$\nabla$ and $\nabla^*$ and mass parameters $m_{\rm crit}$ and $\mu$. 
The Pauli matrix $\tau_3$ acts on a doublet $\chi^\top = (u,d)$ of two light quarks.
Automatic $O(a)$ improvement occurs at maximal twist~\cite{Frezzotti:2003ni}, which here is realized by 
tuning the value of $m_{\rm crit}$ so that the PCAC mass vanishes.
The gauge action is
\begin{equation}\label{TSymaction}
   S^g = \frac{\beta}{3}\sum_x \left(
         \frac{5}{3}  \sum_{\mu,\nu>\mu}\left(1-{\rm Re}\,{\rm Tr}[U_{x,\mu,\nu}^{1\times 1}] \right) 
	-\frac{1}{12} \sum_{\mu,\nu\neq\mu}\left(1-{\rm Re}\,{\rm Tr}[U_{x,\mu,\nu}^{1\times 2}] \right) 
	 \right)\, ,
\end{equation}
where $U_{x,\mu,\nu}^{1\times 1}$ denotes the plaquette term, $U_{x,\mu,\nu}^{1\times 2}$ the
rectangular $1\times 2$ Wilson loops and $\beta$ is the inverse bare coupling. 
Further details concerning the simulation and in particular the tuning to maximal twist can
be found in Ref.~\cite{Boucaud:2008xu}.

\subsection{Correlation functions}
On the lattice the form factors are extracted from dimensionless ratios of
correlation functions. We measure the two- and three- point functions
\begin{eqnarray}
   G(\vec q, t)                  &=&\sum_{\vec x_f} \, e^{-i\vec x_f \cdot \vec q}\, 
                                    \Gamma^0_{\beta\alpha}\, \langle J_{\alpha}(t_f,\vec x_f) \overline{J}_{\beta}(0) \rangle\, , \\
     G_\mu(\Gamma^\nu,\vec q, t) &=&\sum_{\vec x, \vec x_f} \, e^{i\vec x \cdot \vec q}\, 
                                    \Gamma^\nu_{\beta\alpha}\, \langle J_{\alpha}(t_f,\vec x_f)  \Op_\mu(t,\vec x) \overline{J}_{\beta}(0) \rangle \, ,
\end{eqnarray}
where $ \Gamma^\nu = 1/4 [\eins + \gamma_0]$ if $\nu=0$ and 
         $ \Gamma^\nu = 1/4 [\eins + \gamma_0]\gamma_5 \gamma_\nu$ if $ \nu = 1,2 \ {\rm or}\ 3$.
and $J$ is the proton interpolating field. In tmQCD at maximal twist the 
standard interpolating field reads
\begin{equation}\label{interpolatingfield}
   J(x) = \frac{1}{\sqrt{2}}[\eins + i\gamma_5]\epsilon^{abc} \left[ \tilde{u}^{a \top}(x) \C\gamma_5 \tilde{d}^b(x)\right] \tilde{u}^c(x)\, ,
\end{equation}
with $\C$ denoting the charge conjugation matrix.
To enhance the overlap of $J$ with the proton ground state, the quark fields entering
Eq.~(\ref{interpolatingfield}) are smeared,
\begin{eqnarray*}
      \tilde{u}^a(t,\vec x) &=& \sum_{\vec y} F^{ab}(\vec x,\vec y;U(t))\ u^b(t,\vec y) \, ,\\
      F = (\eins + \alpha H)^{N}\, , \quad 
      && \quad H(\vec x,\vec y; U(t)) = \sum_{i=1}^3\left(U_i(x) \delta_{x,y-\hat\imath} + U_i^\dagger(x-\hat\imath) \delta_{x,y+\hat\imath}\right) \, .
\end{eqnarray*}
In addition, the spatial gauge links entering the hopping matrix $H$ are APE-smeared.
Good choices for the parameters $\alpha$ and $N$ were determined in Ref.~\cite{Alexandrou:2008tn}.

In this work,  we restrict ourselves to the axial-vector  and vector currents 
for the operators $\Op_\mu$ inserted into the three point functions.
We use the local currents 
\begin{eqnarray}
   A_\mu(x) &=& \bar u \gamma_\mu \gamma_5 u - \bar d \gamma_\mu \gamma_5 d \\
   V_\mu(x) &=& \bar u \gamma_\mu          u - \bar d \gamma_\mu \gamma_5 d \, ,
\end{eqnarray}
as well as the symmetrized, conserved vector current
\begin{eqnarray}
   V^N_\mu(x) &=& \frac{1}{2} \bigl[j^u_\mu(x)+j^u_\mu(x-\hat \mu)\bigr] - \frac{1}{2} \bigl[j^d_\mu(x)+j^d_\mu(x-\hat \mu)\bigr] \hspace*{0.5cm}{\rm where}\\
   j^q_\mu(x) &=& \bar q(x+\hat \mu) \frac{1}{2}[\gamma_\mu + \eins] U^{-1}_\mu(x) q(x) 
              \ + \ \bar q(x)\frac{1}{2}[\gamma_\mu - \eins] U_\mu(x) q(x+\hat\mu) \, .
\end{eqnarray}

While the two-point functions can be calculated using a forward propagator
from a fixed source, the evaluation of  
three-point functions is more involved.
In correlators containing the isovector operators, disconnected diagrams
are zero up to lattice artifacts, and can be safely neglected
as we approach the continuum limit.
The connected diagrams are calculated 
using sequential inversions through the sink~\cite{Dolgov:2002zm}.
This means that at a fixed source-sink separation we are able to 
obtain results  for all possible 
momentum transfers and insertion times as well as
for any  operator $\Op_\mu$, with a two
sequential inversions per choice of the sink. In this work we use two different
sinks, one optimized for the electric and one for the magnetic form
factor~\cite{Alexandrou:2006ru}. The latter is also suitable for the axial
form factors.

\subsection{Ratios}
In ratios of three- and two- point functions normalization factors and
the leading exponential dependencies on the insertion time cancel.
 From fits to the 
resulting plateaus, the form factors can be extracted. 
Different ratios
can be considered and we here give two examples:
\begin{equation}\label{ratio2}
   R^{\mu}(\Gamma,\vec q,t) = \frac{G^\mu(\Gamma,\vec q,t) }{\sqrt{G(\vec 0, 2(t_f-t))G(\vec q, 2(t-t_i))}}
\end{equation}
or
\begin{equation}\label{ratio1}
   R^{\mu}(\Gamma,\vec q,t) = \frac{G^\mu(\Gamma,\vec q,t) }{G(\vec 0, t_f)}\ \sqrt{\frac{G(\vec q, t_f-t)G(\vec 0,  t)G(\vec0,   t_f)}
                                                                                   {G(\vec 0  , t_f-t)G(\vec q,t)G(\vec q,t_f)}} \, .
\end{equation}
 \begin{figure}[h]
\includegraphics[width=0.47\linewidth, height=0.34\linewidth]{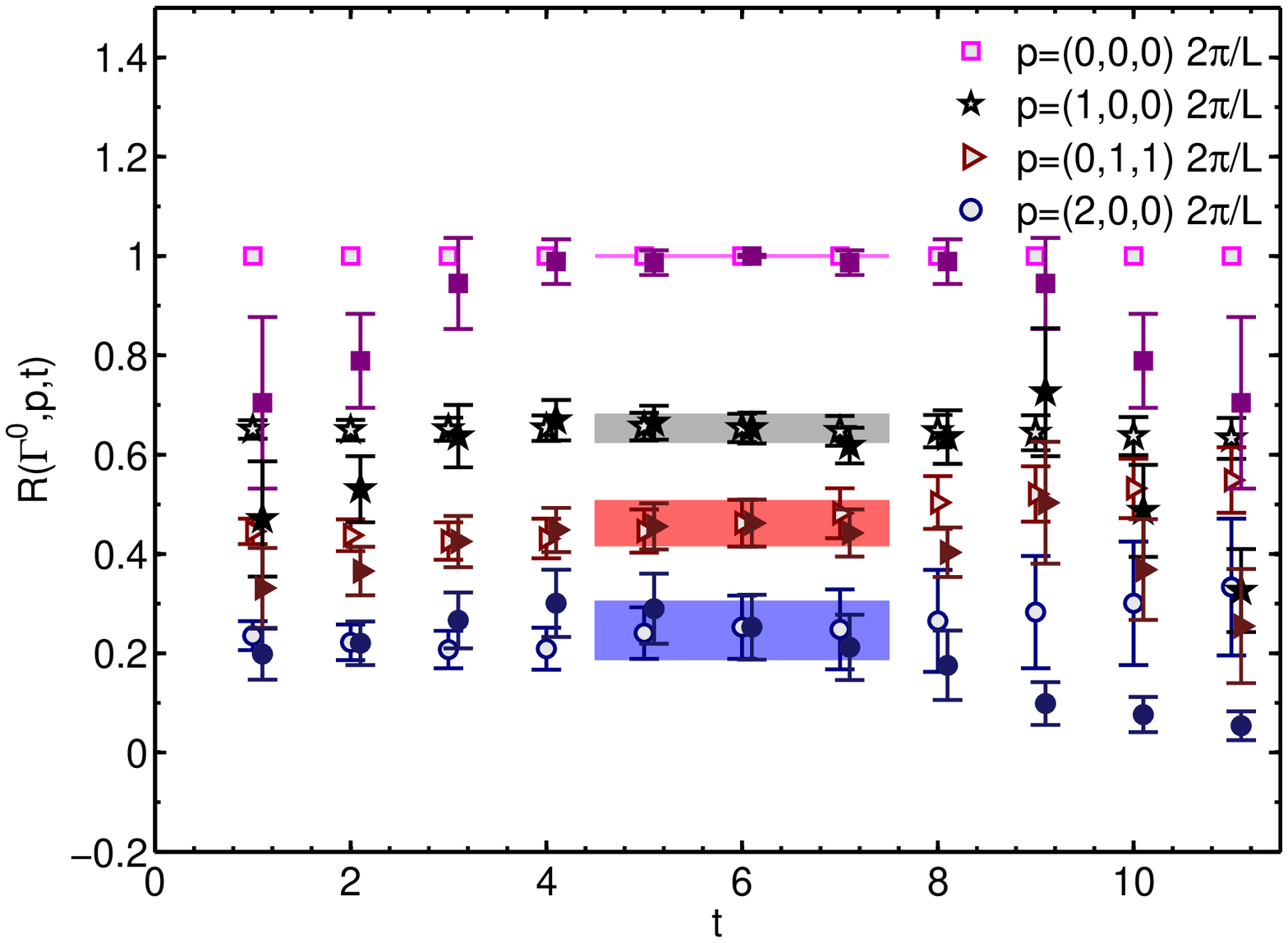}\hspace{0.9cm}
\includegraphics[width=0.47\linewidth, height=0.34\linewidth]{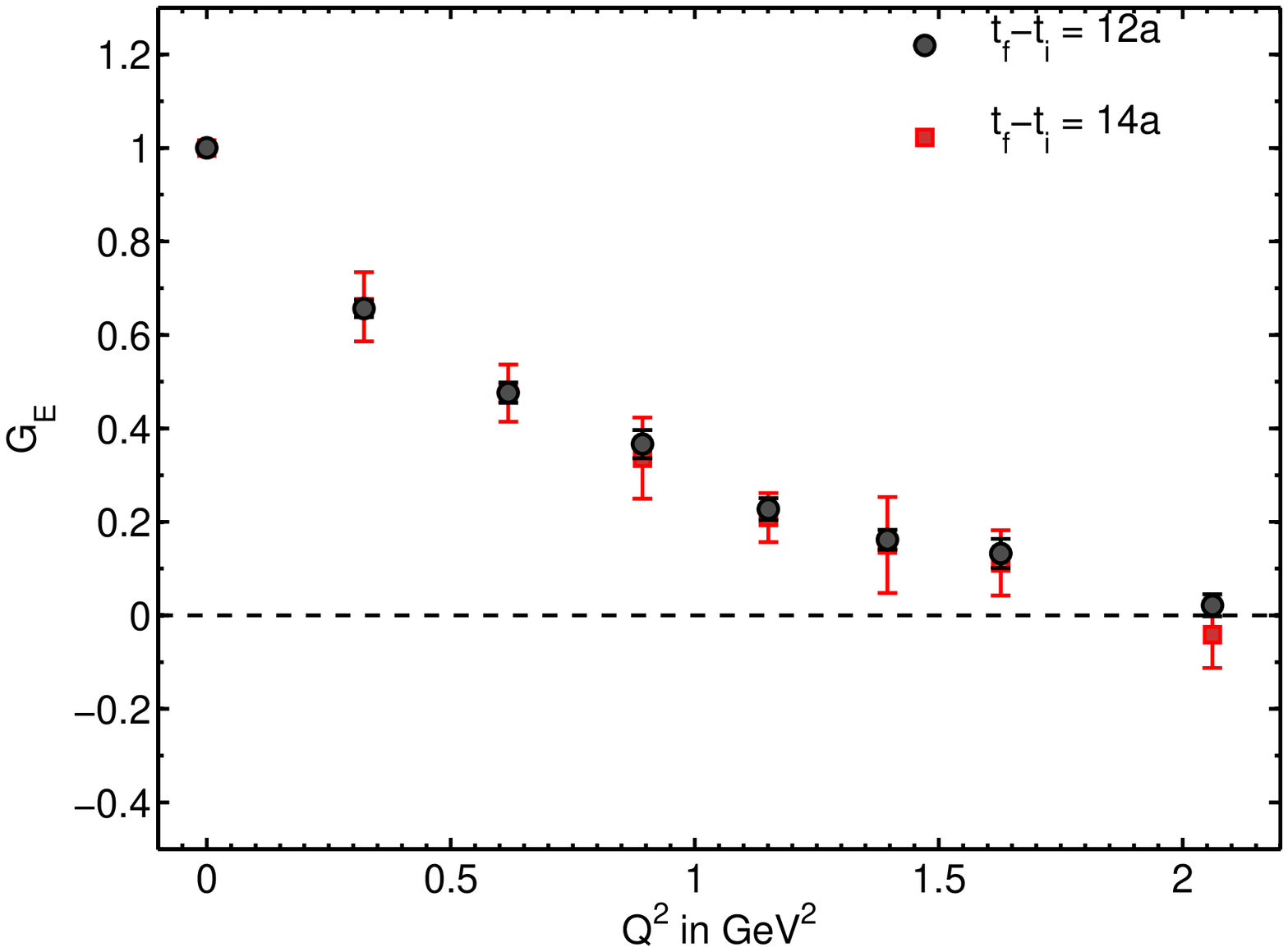}
\caption{Left panel: Comparison of the ratios given in Eqs.~(2.10)  (filled symbols) and (2.11) (open symbols)
         for source type given in Eq.~(2.13) and a few representative momentum combinations.
         Right panel: Electric Sachs form factor extracted with a sink source separation of $12a$ and 
	 $14a$.}
\label{ratios_Fig}
\end{figure}
As shown in Fig.~\ref{ratios_Fig}, both lead to compatible plateaus
but to different statistical errors. 
We use the ratio of Eq.~(\ref{ratio1}) for the final
analysis, which turns out to be superior because it does not contain potentially noisy two point functions at large
separations and because 
correlations between its different factors reduce the statistical noise.
This being most evident at zero momentum with the conserved current, where 
noise cancels completely.

Once the plateau values
\begin{equation}
\Pi(\Gamma,\vec q) = \lim_{t_f-t \to \infty}\ \lim_{t-t_i \to \infty} R(\Gamma,\vec q,t) 
\end{equation}
are estimated, the form factors are obtained from the following 
combinations
\begin{eqnarray}
   \Pi^\mu(\Gamma^0,\vec q) &=& \frac{c}{2m}[(m+E)\delta_{0,\mu} + \sum_k i q_k \delta_{k,\mu}]\ \, G_E(Q^2)  \label{Etype} \\
   \Pi^i(\Gamma^1,\vec q) + \Pi^i(\Gamma^2,\vec q) + \Pi^i(\Gamma^3,\vec q) &=& \frac{c}{2m}\sum_{jkl}\epsilon_{jkl} q_j \delta_{l,i} \ G_M(Q^2) \\
   \Pi^{5i}(\Gamma^1,\vec q) + \Pi^{5i}(\Gamma^2,\vec q) + \Pi^{5i}(\Gamma^3,\vec q) &=& \frac{ic}{4m}[(q_1+q_2+q_3)\frac{q_i}{2m}\ G_p(Q^2)-(E+m)\ G_A(Q^2)] \, .\qquad
\end{eqnarray}
$E$ is the energy of a proton with momentum $\vec p_i$, $m$ its mass and the constant
$c=\sqrt{\frac{2m^2}{E(E+m)}}$ arises from the normalization of lattice hadron states
that we use.
Note that the sequential source needed in the calculations of the magnetic Sachs 
form factor is the same as the one for $G_A$ and $G_p$.

\section{Results}
We perform the calculation at three pion masses with a fixed 
lattice spacing. The source and sink time slices are taken  at
 $t_i=0$ and $t_f=12$ and a
check with $t_f-t_i=14$ on 96 configurations at the highest mass gave compatible results, 
as shown in the right panel of Fig.~\ref{ratios_Fig}. The 
remaining  parameters of the calculation together
 with the most important results
are summarized in Table.~\ref{tab1}.\begin{table}[h]
\centering
\begin{tabular}{c c c c c c c}
\toprule
      $m_\pi [GeV]   $& number of confs &$ m_{N}\ [GeV] $&$ \sqrt{\langle r_1^2 \rangle}\ [fm]$&$ \sqrt{\langle r_2^2 \rangle}\ [fm] $&$ \mu_{IV} \ [\mu_N] $&$ g_A      $\\
     \midrule
      $0.4470(12)    $&  346   &$ 1.287(13)            $&$ 0.489(25)                         $&$ 0.558(61)                          $&$ 2.83(21)           $&$ 1.171(41)$\\
      $0.3903(9)     $&  184   &$ 1.245(9)             $&$ 0.527(34)                         $&$ 0.607(79)                          $&$ 2.90(36)           $&$ 1.096(47)$\\
      $0.3131(16)    $&  419   &$ 1.143(11)            $&$ 0.649(38)                         $&$ 0.63(11)                           $&$ 2.85(46)           $&$ 1.23(10) $\\
     \bottomrule
\end{tabular}
\caption{The form factors are extracted using a lattice
of size  $24^3\times 48$ at $\beta=3.9$ with a lattice spacing of
          $a=0.0889(12)$. This table summarizes the main results at the three different pion masses.}\label{tab1}
\end{table}

\subsection{Electric and magnetic Sachs form factors}
\begin{figure}[h]
\includegraphics[width=0.47\linewidth, height=0.34\linewidth]{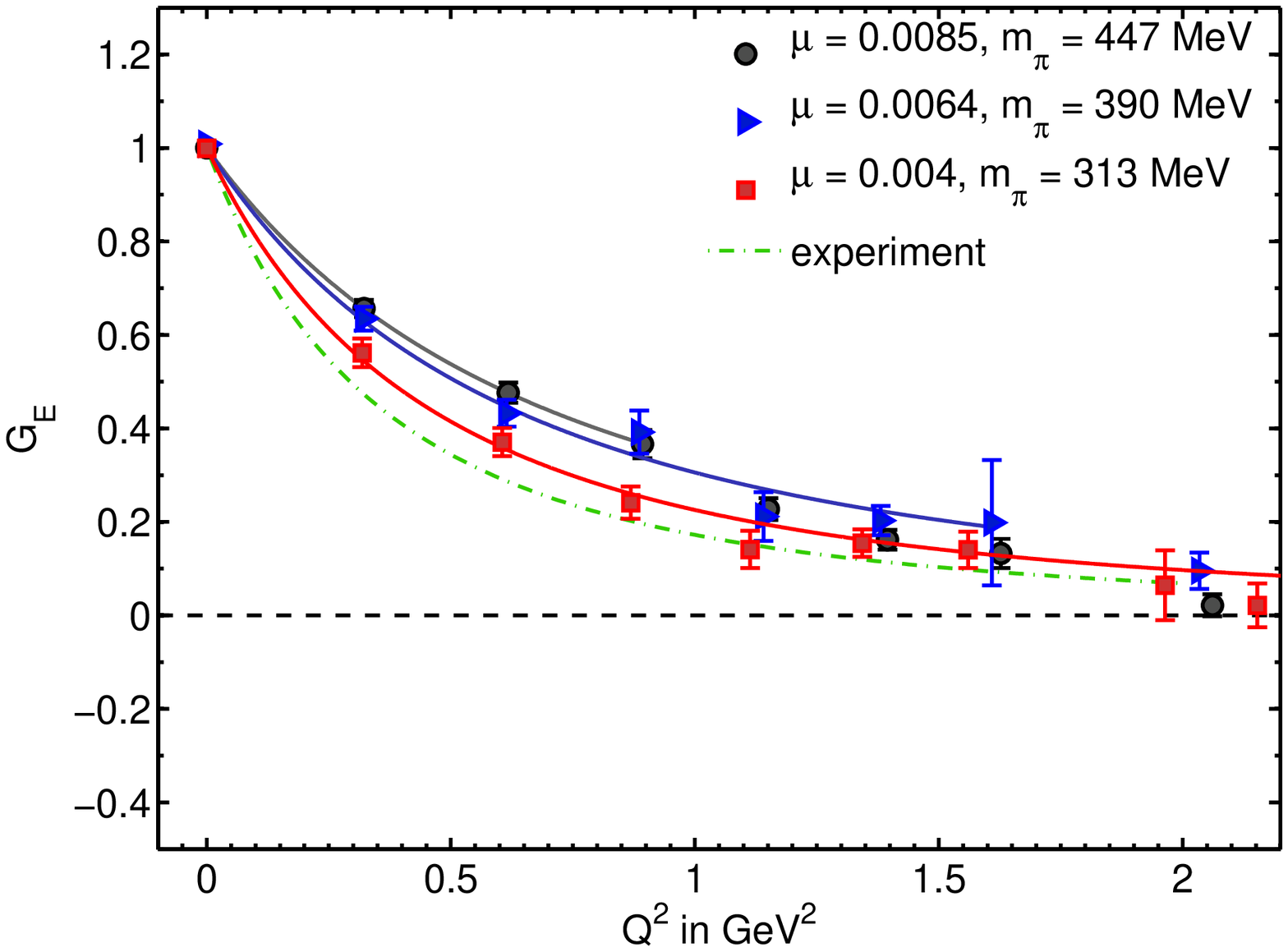}\hspace{0.9cm}
\includegraphics[width=0.47\linewidth, height=0.34\linewidth]{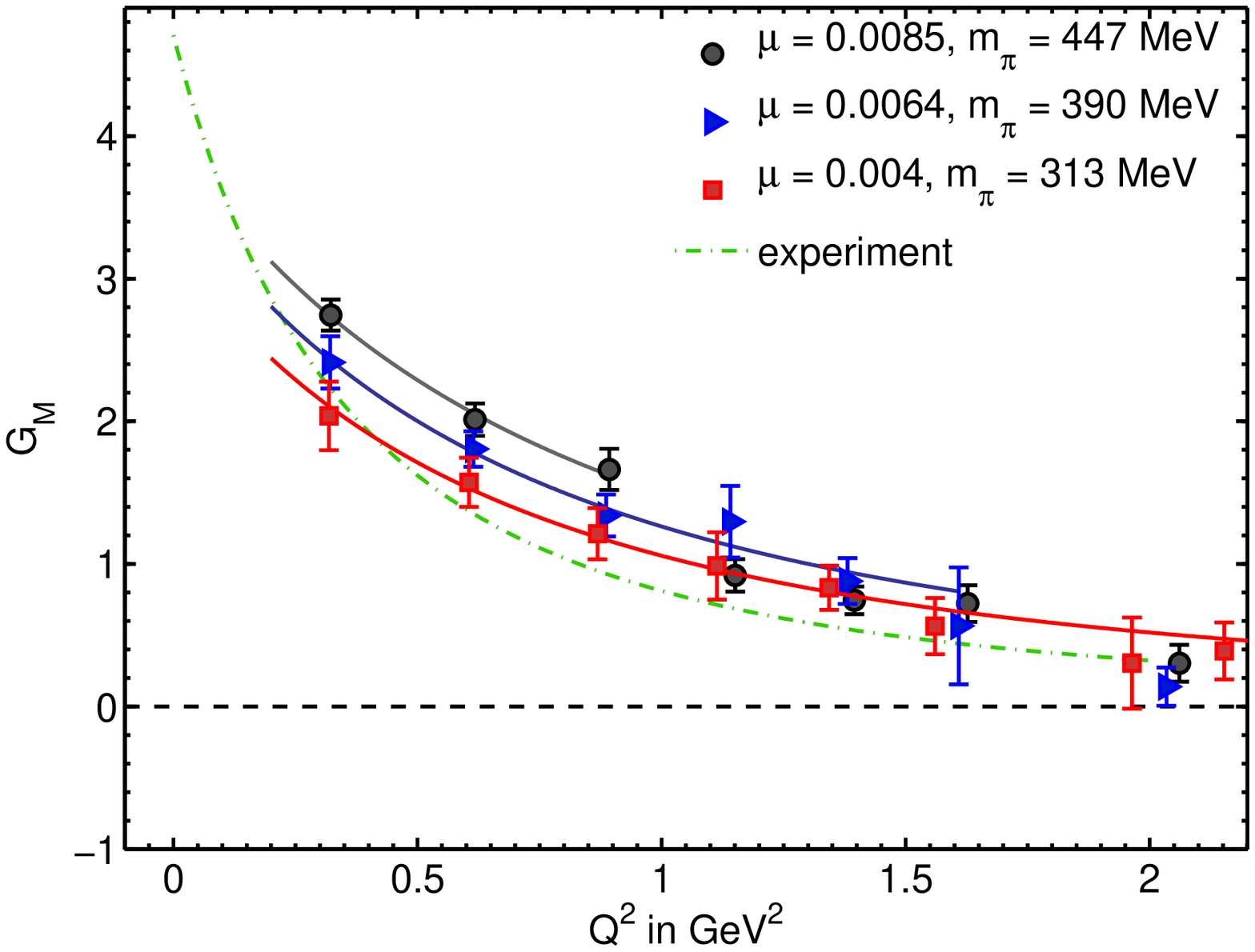}
\caption{$Q^2$-dependence of $G_E$ (left) and $G_M$ (right). The dashed curve is
a dipole fit to experimental data.}\label{gEgM_Fig}
\end{figure}

The dependence of
$G_E$ and $G_M$ on the euclidean momentum transfer squared is shown
in Fig.~\ref{gEgM_Fig}. 
A dipole form 
\begin{equation}
   G_E(Q^2) =      \frac{1}{\left(1+\frac{Q^2}{m_E^2}\right)^2} \, , \qquad \qquad
   G_M(Q^2) = \frac{G_M(0)}{\left(1+\frac{Q^2}{m_M^2}\right)^2}   
\end{equation}
describes our data very well and the solid lines in Fig.~\ref{gEgM_Fig}
are the corresponding least squares fits. From the slope at
zero momentum transfer, an electric and magnetic radius can be defined
\begin{eqnarray} 
   \langle r^2_{E,M} \rangle = -\frac{6}{G_{E,M}(Q^2)}\ \frac{d\, G_{E,M}(Q^2)}{dQ^2}  \biggr|_{Q^2=0}  = \frac{12}{m^2_{E,M}}\, .
\end{eqnarray}
These can easily be translated into the more common Dirac and Pauli radii
\begin{equation}
   \langle r_1^2\rangle = \frac{12}{m_E^2} - \frac{3G_M(0)-3}{2m^2} \, , \qquad \qquad
   \langle r_2^2\rangle = \frac{6}{G_M(0)-1} \left(\frac{2G_M(0)}{m_M^2} - \frac{2}{m_E^2} \right) + \frac{3}{2m^2} \, .
\end{equation}
\begin{figure}[h]
\includegraphics[width=0.47\linewidth, height=0.34\linewidth]{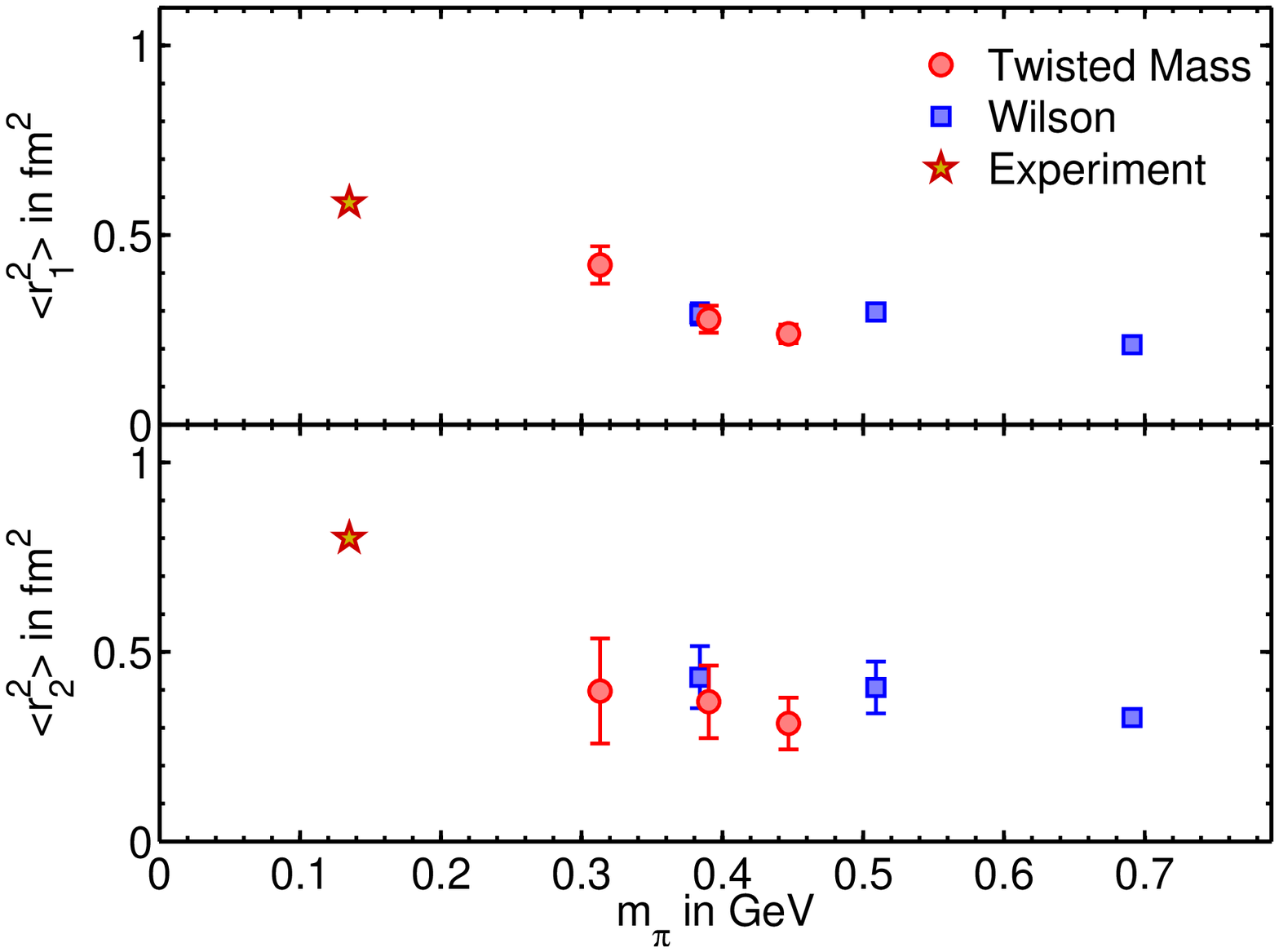}\hspace{0.9cm}\vspace*{-0.2cm}
\includegraphics[width=0.47\linewidth, height=0.34\linewidth]{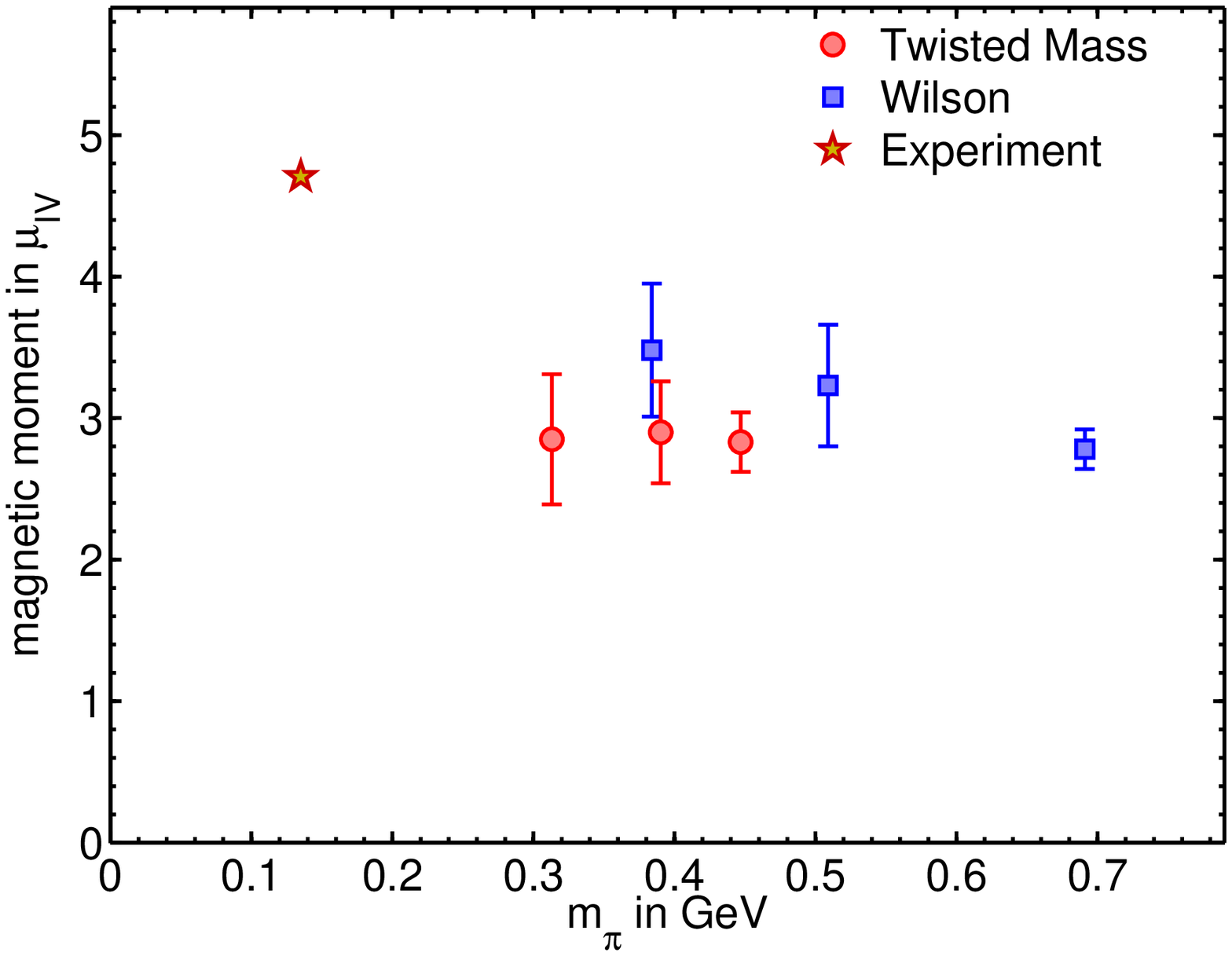}
\caption{Comparison of results on $\langle r_1^2 \rangle$, $\langle r_2^2\rangle$ and $\mu_{IV}$ 
between $N_F=2$ twisted mass and Wilson fermions.}\label{chiPT_Fig}
\vspace*{-0.8cm}
\end{figure}

\begin{figure}[h]
\includegraphics[width=0.47\linewidth, height=0.325\linewidth]{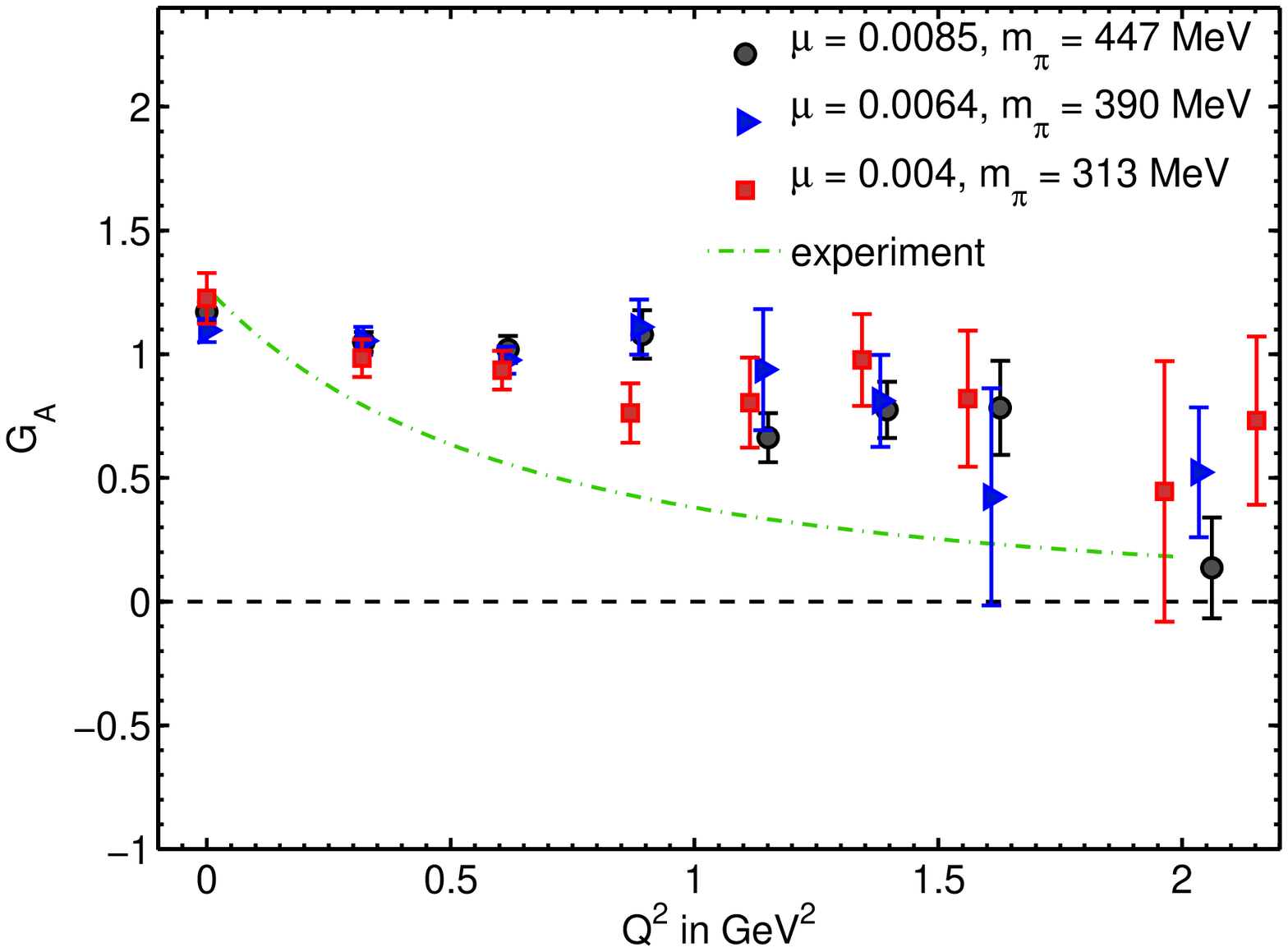}\hspace{0.9cm}
\includegraphics[width=0.47\linewidth, height=0.325\linewidth]{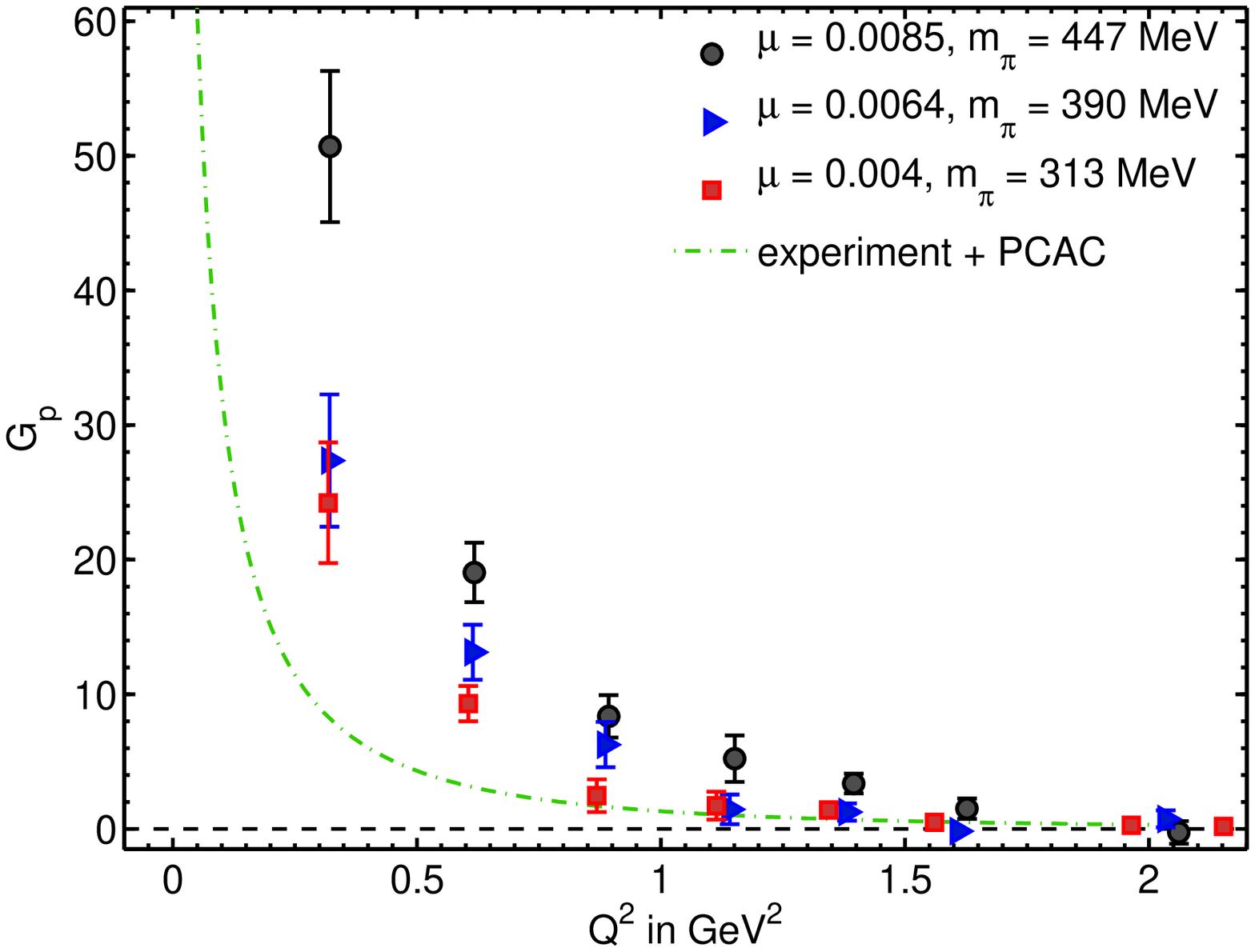}
\caption{$Q^2$-dependence of $G_A$ (left) and $G_p$ (right). The dashed
 green curve (left) is a dipole fit to experimental data on $G_A$ yielding $m_A=1.1$~GeV.
In the right figure it derives from
the fit to experimental results on $G_A$ together with the assumption of
pion pole dominance.}\label{gAgp_Fig}
\end{figure}

Using the value of the lattice spacing determined from an
analysis of  the nucleon mass within the twisted mass QCD framework 
we convert our lattice quantities to physical units and give the results in Table~\ref{tab1}.
The magnetic form factor at zero momentum transfer is directly related to 
the magnetic moment, $G_M(0) \frac{e\hbar}{2m_N} = \mu_{IV}$. 
In Table~\ref{tab1} 
we give the isovector magnetic moment converted to nuclear magnetons
using the physical nucleon mass.
We show the dependence of $\langle r_1^2 \rangle$, $\langle r_2^2\rangle$ and $\mu_{IV}$ on
the pion mass in Fig.~\ref{chiPT_Fig}.  For comparison we include
 the corresponding results of 
Ref.~\cite{Alexandrou:2006ru}, which were obtained with two degenerate
 dynamical Wilson
fermions in a similar setup. 
As can be seen, the results in the two formulations are in good agreement.
With the isovector current, the corresponding 
physical observable is the magnetic moment of the proton minus the one of the
neutron. This is the experimental point
 shown in the right panel of Fig.~\ref{chiPT_Fig}.
As can be seen, the experimental values in all cases are higher than lattice
results.
A chiral extrapolation
of our data to the physical point will be carried out once
we obtain results at an  additional lighter pion mass. This additional
input is needed in order to  obtain
reliable results at the physical point.

\subsection{Axial form factors}
The axial form factors are shown in
Fig.~\ref{gAgp_Fig} with  $G_p$ showing the larger statistical errors.
We get a good signal for the axial charge $g_A = Z_A\, G_A(0)$, with
the renormalization constant $Z_A=0.76(1)$ that has been computed in~\cite{Dimopoulos:2007fn}.
We give our values in Table.~\ref{tab1}.

\vspace{-0.4cm}
\section{Conclusions}
\vspace{-0.15cm}
The isovector electromagnetic and axial nucleon form factors are 
evaluated using two dynamical degenerate twisted mass fermions for pion
masses down to about 300 MeV. 
The results are in agreement with previous lattice studies~\cite{Alexandrou:2006ru}. 
Like in Ref.~\cite{Alexandrou:2006ru}, we
find better agreement with experiment in the case of the electric form factor
than for the magnetic one. Although our results on the
 axial charge are in good agreement with 
experiment, we find a weaker  momentum dependence for $G_A$. At low
$Q^2$,  $G_p$ shows a steep increase as expected from pion pole dominance and
approaches the theoretical prediction as the pion mass decreases.
In the future we plan to analyze configurations at a lighter pion
mass and larger volume. This will enable us to reliably carry out
a chiral extrapolation of the magnetic moment, radii and axial charge
to the physical point.

\vspace{-0.25cm}
\bibliographystyle{h-elsevier}
\bibliography{ffrefs}

\end{document}